\documentclass[apj]{emulateapj}
\usepackage{mathptmx}
\newcommand  \kms      {\ifmmode {\rm km\,s}^{-1} \else km\,s$^{-1}$\fi}

\newcommand  \ergs     {\ifmmode {\rm ergs\,s}^{-1} \else ergs s$^{-1}$\fi}
\newcommand  \ergcms   {\ifmmode {\rm ergs\,cm}^{-2}\,{\rm s}^{-1}
                        \else ergs\,cm$^{-2}$\,s$^{-1}$\fi}
\newcommand  \ergcmsA {\ifmmode{\rm ergs\,cm}^{-2}\,{\rm s}^{-1}\,{\rm\AA}^{-1}
                        \else ergs\,cm$^{-2}$\,s$^{-1}$\,\AA$^{-1}$\fi}
\newcommand \ergcmsHz {\ifmmode{\rm ergs\,cm}^{-2}\,{\rm s}^{-1}\,{\rm Hz}^{-1}
                        \else ergs\,cm$^{-2}$\,s$^{-1}$\,Hz$^{-1}$\fi}
\newcommand  \phcms    {\ifmmode {\rm ph\,cm}^{-2}\,{\rm s}^{-1}
                        \else ,ph\,cm$^{-2}$\,s$^{-1}$\fi}
\newcommand  \phcmsA   {\ifmmode {\rm ph\,cm}^{-2}\,{\rm s}^{-1}\,{\rm\AA}^{-1}
                        \else ph\,cm$^{-2}$\,s$^{-1}$\,\AA$^{-1}$\fi}
\def\micron{\ifmmode \mu{\rm m} \else $\mu$m\fi}
\def\kms{\ifmmode {\rm km\,s}^{-1} \else km\,s$^{-1}$\fi}
\def\Hubble{\ifmmode {\rm km\,s}^{-1}\,{\rm Mpc}^{-1}
        \else km\,s$^{-1}$\,Mpc$^{-1}$\fi}
\def\ergsec{\ifmmode {\rm ergs\;s}^{-1} \else ergs s$^{-1}$\fi}
\def\ergscm{\ifmmode {\rm ergs\,s}^{-1}\,{\rm cm}^{-2}
          \else ergs\,s$^{-1}$\,cm$^{-2}$\fi}
\def\ergscmA{\ifmmode {\rm ergs\,s}^{-1}\,{\rm cm}^{-2}\,{\rm \AA}^{-1}
          \else ergs\,s$^{-1}$\,cm$^{-2}$\,\AA$^{-1}$\fi}
\def\ergscmHz{\ifmmode {\rm ergs\,s}^{-1}\,{\rm cm}^{-2}\,{\rm Hz}^{-1}
          \else ergs\,s$^{-1}$\,cm$^{-2}$\,Hz$^{-1}$\fi}
\def\Msun{\ifmmode M_{\odot} \else $M_{\odot}$\fi}
\def\Lsun{\ifmmode L_{\odot} \else $L_{\odot}$\fi}
\def\qo{\ifmmode q_{0} \else $q_{0}$\fi}
\def\Ho{\ifmmode H_{0} \else $H_{0}$\fi}
\def\ho{\ifmmode h_{0} \else $h_{0}$\fi}
\def\qo{\ifmmode q_{0} \else $q_{0}$\fi}
\def\ao{\ifmmode a_{0} \else $a_{0}$\fi}
\def\to{\ifmmode t_{0} \else $t_{0}$\fi}
\def\ltsim{\raisebox{-.5ex}{$\;\stackrel{<}{\sim}\;$}}
\def\gtsim{\raisebox{-.5ex}{$\;\stackrel{>}{\sim}\;$}}
\def\Halpha{\ifmmode {\rm H}\alpha \else H$\alpha$\fi}
\def\Hbeta{\ifmmode {\rm H}\beta \else H$\beta$\fi}
\def\hb{\ifmmode {\rm H}\beta \else H$\beta$\fi}
\def\Hgamma{\ifmmode {\rm H}\gamma \else H$\gamma$\fi}
\def\Hdelta{\ifmmode {\rm H}\delta \else H$\delta$\fi}
\def\Lya{\ifmmode {\rm Ly}\alpha \else Ly$\alpha$\fi}
\def\Lyb{\ifmmode {\rm Ly}\beta \else Ly$\beta$\fi}
\def\hi{\ifmmode \mbox{{\rm H}\,{\sc i}} \else H\,{\sc i}\fi}

\def\ciii{\ifmmode {\rm C}\,{\sc iii} \else C\,{\sc iii}\fi}

\def\o5007{[O\,{\sc iii}]\,$\lambda5007$}

\def\ne212m {[Ne\,{\sc ii}]\,$12.8 \mu m$}

\def \Ledd{$L/L_{Edd}$}

\def  \kms         {\hbox{km s$^{-1}$}}          
\def  \ergs        {\hbox{erg s$^{-1}$}}              

\def  \La          {\ifmmode {\rm Ly}\alpha \else Ly$\alpha$\fi}
\def  \Ka          {\ifmmode {\rm K}\alpha \else K$\alpha$\fi}
\def  \Lb          {\ifmmode {\rm L}\beta \else L$\beta$\fi}
\def  \Ha          {\ifmmode {\rm H}\alpha \else H$\alpha$\fi}
\def  \Hb          {\ifmmode {\rm H}\beta \else H$\beta$\fi}
\def  \Pa          {\ifmmode {\rm P}\alpha \else P$\alpha$\fi}
\def  \CIIIb       {\ifmmode {\rm C}\,{\sc iii]}\,\lambda1909
                     \else C\,{\sc iii]}\,$\lambda1909$\fi}
\def  \CIV         {\ifmmode {\rm C}\,{\sc iv}\,\lambda1549
                     \else C\,{\sc iv}\,$\lambda1549$\fi}
\def  \MgII         {\ifmmode {\rm Mg}\,{\sc ii}\,\lambda2798
                     \else Mg\,{\sc ii}\,$\lambda2798$\fi}
\def  \OVI         {\ifmmode {\rm O}\,{\sc vi}\,\lambda1035
x
                     \else O\,{\sc vi}\,$\lambda1035$\fi}
\def \chandra  {{\it Chandra}}

\def \GR       {GRO J1655--40}

\shorttitle{Thermal wind in GRO J1655-40}
\shortauthors{Netzer, H}

\begin{document}

\title{A Thermal Wind Model for the X-ray Outflow in \GR}

\author{
Hagai Netzer,\altaffilmark{1,2}
}

\altaffiltext{1}
          {School of Physics and Astronomy and the Wise
                Observatory, The Raymond and Beverly Sackler Faculty of
                Exact Sciences, Tel-Aviv University, Tel-Aviv 69978,
                Israel}
\altaffiltext{2}
 {Kavli Institute for Particle Astrophysics and Cosmology and Department of Physics, Stanford University, CA 94305}

\begin{abstract}
Recent \chandra\ observations of an outflowing gas in \GR\ resulted in a suggestion by Miller et al. (2006)
that the wind in this system must be powered by a magnetic process that can also drive accretion through
the disk around the black hole. The alternative explanations, of radiation
pressure or thermally driven flows, were considered unsatisfactory because of the highly ionized level of the gas
and because of the derived small distance from the black hole, well inside the minimum distance required
for an efficient X-ray heated wind.
The present paper shows that there is a simple photoionized wind solution for this system where
the gas is much further out than assumed by Miller et al., at $r/r_g = 10^{4.7-5.7}$.
The expected wind velocity, as well as the computed
equivalent widths of more than 50 absorption
lines in this single-component 1D model, are all in good agreement with the \chandra\ observations.

\end{abstract}

\keywords{
binaries: accretion, accretion disks - spectroscopic - black hole physics - X-rays: stars - stars: individual (GRO J1655-40)
}

\section{Introduction}
High resolution spectroscopy of transient X-ray sources is a powerful tool for studying
the nature of such systems.
The recent paper by Miller et al. (2006, hereafter M06) is an excellent example that illustrates the power
of such methods. These authors analyzed the high energy transmission grating
(HETG) \chandra\ observations of \GR\ (Orosz \& Baily 1997),
a stellar-mass binary system containing a 7 \Msun\  BH and a F3IV-F6IV companion
with a mass of 2.3 \Msun\ at a mean distance of about $10^6$ gravitational radii ($r_g=GM/c^2$).
The observations were
obtained during an X-ray bright phase when the intrinsic luminosity of the source
approached $3.3 \times 10^{37}$ \ergs, equivalent to \Ledd$\sim 0.03$.
They show the clear signature of an outflow (``X-ray wind'').
M06 measured the equivalent widths (EWs) and the line widths of more than 70 identifiable
X-ray lines. Using empirical fits to the
line profile, they derived  column densities for the
more abundant ions (mostly H-like
and He-like lines of oxygen, neon, magnesium, silicon, sulfur, argon, calcium, iron and nickel)
and obtained reliable estimates of the gas composition and level of ionization. Fitting the
unblended lines indicate outflow velocities of 300--1000 \kms\
and line widths of  300--500 \kms.
This data set is not the first of its kind (e.g. Miller et al. 2004) but, so far, it is the highest quality one.

According to M06, the X-ray wind in \GR\ must be powered by magnetic processes (pressure generated
by magnetic viscosity  or magnetocentrifugal forces, e.g. Blandford and Payne 1982).
The reason for favoring this interpretation is the inefficiency of
two alternative mechanisms, radiation pressure force and thermal expansion, to explain the observed  properties.
 This conclusion, claimed to be the best evidence so far for a magnetically driven disk wind,
 was based on a detailed photoionization model for the X-ray heated gas.
The outflowing gas in this model is
very close to the BH, some $450 \, r_g$ from it, where the escape velocity is about 2$\times 10^4$ \kms.
This location was inferred from trying to match the observed EWs of more than
70 absorption lines. The best fit model suggests a very high ionization parameter, gas
density of $n \simeq 5.6 \times 10^{15}$ cm$^{-3}$ and a column density  of $N_H \simeq 10^{24.15}$ cm$^{-2}$.
The observed wind velocity is more than an order of magnitude smaller
than the escape velocity at the suggested location. This was explained to be due to the high inclination
of the disk (67--85 degrees) in this system. The highest velocity component is perpendicular to the accretion
disk and the observed velocities are much smaller.
Given this location and level of ionization,
 the radiation pressure force is too small to drive the  wind.
A thermal wind solution was rejected on grounds that the gas location is orders of magnitude inside the region
where such winds are likely to develop (e.g. Begelman, McKee and Shields 1983). The rejection of these two mechanisms
leave magnetic processes as the only viable explanation for the outflow.

In this paper I show that there is, in fact, a simple wind solution that is consistent with
all the observed properties of this system. \S2 presents this solution and \S3 discusses several other implications.

\section{A thermal wind model for GRO J1655--40}
\subsection{General considerations}

Thermal winds are expected to arise in various astrophysical situations involving X-ray irradiated accretion disks.
These includes disks around galactic X-ray sources as well as the inner regions of active galactic nuclei (AGNs).
In such cases, the central X-ray source provides the energy to lift the gas from the surface of the
disk at large distances and to heat it to a high enough
temperature that enables escape via an expanding wind.
Such winds have been discussed extensively in the literature. Here I
focus on the Begelman et al. (1983) analytical approach and the follow-up 2D time-dependent numerical calculations of
Woods et al. (1996, hereafter W96).

Important parameters in X-ray irradiated thermal wind models are the Compton temperature of the gas,
$T_C$, and the Compton radius, $r_C$, which is the distance at which the escape velocity equals the isothermal
sound speed at the Compton temperature. This radius is given by
\begin{equation}
r_C = \frac{9.8 \times 10^9}{T_{C8} } \frac{M(BH)}{M_{\odot}} \,\, {\rm cm} \,\, ,
\end{equation}
where $T_{C8}=T_C/10^8 \, {\rm K}$. According to the theoretical estimates, X-ray heated winds can develop at
any $r \gtsim 0.1 r_C$. The more accurate numerical calculations (e.g. Fig. 4 in W96) give even
smaller radii, as small as $0.01 r_C$.
There are various wind regions depending on the luminosity
(isothermal wind, steadily heated free wind and gravity inhibited wind, see W96
for more details).
The Compton temperature of the gas depends on the spectral energy distribution (SED)
of the continuum source.
In the case of \GR, this is obtained directly from the observations, after correcting for galactic
absorption. There is little uncertainty in the chosen SED since most of the energy is emitted over the observed
\chandra\ band. Here  I adopt the SED described in M06 which is a combination of
a thermal source with $kT=1.35$ keV and a high energy power-law of the form
$L_E \propto E^{-2.54}$. This results in
$T_C\simeq 1.4 \times 10^7$ K. Thus,
$r_C \simeq 10^{11.7} $ cm and an effective thermal wind is likely to develop at all distances exceeding
about $10^{10.7}$ cm ($10^{4.7}r_g$) and perhaps much smaller.

The simple 1D model presented here assumes a point-like X-ray source in the center of the disk and a $4 \pi$
ionized gas distribution that may have different density and column density at different polar angles.
This is {\it not} a full numerical model but rather a demonstration that a differentially
expanding, photoionized thermal wind can explain
the \chandra\ observations of \GR. Realistic wind models combine the micro-physics of the ionized gas with the
gas dynamics, in a time dependent fashion. Such calculations are 
are beyond the scope of the present work or the work of M06.
However, as shown by Begelman et al. (1983), many of the wind properties 
 can be inferred from general considerations.
Further confirmation can be found in Chelouche and Netzer (2005) who
describe a phenomenological approach of a polytropic wind that shares many properties with 
numerical 1D and 2D wind simulations. Simple estimates of this type are nevertheless limited in the
ability to estimate the
wind velocity and density at various locations.

Thermal wind are characterized by two regimes, before and after crossing the critical (sonic) point.
The present work addresses the region 
around and beyond the sonic point, where most of the observed line opacity is assumed to take
place. In this region the velocity increases slowly with radius in a way that can be
expressed as
\begin{equation}
v_{wind} \propto r^x \,\, ,
\end{equation}
 where in most cases of interest $0 \le x < 0.5$. The wind speed at  $r \gg r_C$ approaches 3--4 time
 the sound speed in the fully ionized gas.
AS shown by W96, thermal outflows can develop in gas whose temperature is much below $T_C$
(e.g. W96 Fig. 7). Such gas is carried away with the flow and can reach, under some circumstances, its Compton
temperature. The case addressed here is that of the cooler gas although, for simplicity, some of the estimates
are based on the assumption $T_{wind}=T_C$.

Continuity in a spherical geometry (see however comments in \S3) requires that
\begin{equation}
n \propto r^{-(2+x)} \,\, .
\end{equation}
Assign $n_0$ as the gas (hydrogen) density at the base of the wind, just beyond the sonic point,
where the distance from the BH is $r_0$. This gives a maximum column density of
\begin{equation}
N_H =n_0 r_0 /(1+x) \,\, .
\end{equation}
The ionization parameter of the gas is given by
\begin{equation}
\xi = \frac{L_X}{n r^2} \,\, ,
\end{equation}
where  $L_X$ is the integrated luminosity above 1 Rydberg and where
at the base of the wind $\xi_0=L_X/n_0 r_0^2$.
Thus, $\xi \propto r^x$.

Experimenting with the above SED and a detailed photoionization code (see \S2.2), and changing  the
ionization parameter until it results in a fractional 
ionization that fits the observations to within a factor two, I
obtain $\xi \simeq 10^3$. This is 15--50 times smaller than the value suggested by
M06\footnote{The reason for this discrepancy, which is a major
cause for the very different model presented here, is not clear}.
Given the column densities inferred from the EWs of several unsaturated lines (M06
supplementary Table 1), and assuming near solar composition, I derive a minimum required column density
of $N_H \simeq 10^{23.6}$ cm$^{-2}$, about a factor 3 smaller than the column density used in M06.
The combination of $\xi_0$ and $N_H$, together with
$L_X$, can be used to solve for $n_0$ and the maximum value of $r_0$ which is about
$10^{11}$ cm ($\sim 10^5 r_g$). This location is outside the minimum distance
($0.1r_C$ or smaller) required for a thermal wind.
Some 90\% of the wind opacity is achieved over the  distance 1--10$r_0$. For \GR,
this corresponds to $10^{4.7} \ltsim r/r_g \ltsim 10^{5.7}$. These dimensions fit nicely with the expected
dimensions of the accretion disk in this system.
 The expected wind velocities over this region are similar to the velocities
inferred from the observed line blueshifts in the spectrum of the source.

Thus, estimates based on the observed ionization, column density and velocity of the gas in \GR, all agree
with the conditions expected in a thermal wind around an X-ray irradiated accretion disk.

\subsection{A photoionized wind model for \GR}
Detailed photoionization calculations were carried out to compare the predicted
spectra, under such conditions, with the \chandra\ observations.
 The photoionization code used, ION06, is the 2006 version of the code
ION most recently described in Netzer (2004) and in Netzer et al. (2005).
The only significant modification is the introduction of new radiative and di-electronic recombination
rates described in Badnell (2006) and references therein. The latest atomic data for iron L-shell
lines, published by Landi and Gu (2006), were used in order to facilitate a meaningful comparison with the large number of Fe~{\sc xxii}
and Fe~{\sc xxiii} lines observed in the spectrum of this source.
The SED and $L_X$ are those given in M06. The gas composition was assumed to be solar except for
calcium and iron where, following the suggestion of M06, twice the solar abundances were assumed.

Several velocity profiles were tested assuming $0 \le x < 0.5$. Line widths of
250--500 \kms\ were chosen to agree
with the observations. These velocities can be interpret as due to internal (sound or turbulent)
or global (wind expansion) motion.
The results are not sensitive to
the exact values of $\xi$, $r_0$ and $n_0$ provided they are within a factor $\sim 2$ of the values given above.
Several models with $\xi$ decreasing with $r$ have also been computed and found to be of similar quality.

Two additional parameters require further explanation. The first is the covering fraction of the outflowing gas which,
given the disk-wind geometry, may not be the same for the emission and absorption lines. As noted by M06, the
very weak emission lines and the lack of clear P-Cygni profiles suggest a small emission
measure of the X-ray gas. This can be interpreted as different density
and/or column density at different polar angles. For example, the W96 numerical calculations
(Fig. 7 and Tables 2--4) clearly
show that column densities along the plane of the disk can be significantly larger than the column perpendicular
to the disk surface. Such a scenario will produce weak emission lines. In the 1D photoionization calculations
presented here, this
is taken into account by assuming a small emission (global) covering factor, $\Omega_{em}/4 \pi \le 0.2$.

There are indications in the
spectrum that the absorption (line-of-sight) covering factor is also smaller than unity.
This can be the result of a patchy
absorber or due to scattering of the X-ray continuum into the line of sight.
For example, several of the longer wavelength lines,
e.g. those of neon and magnesium, are resolved yet the optical depths inferred from the
profiles of the
low and the high series lines are inconsistent and suggest an unobscured continuum. A second example is
the pair of Fe~{\sc xxiv} lines at 10.67,10.62\AA. The optical depths of the two differ by a factor 2 yet
the observations suggest almost identical depths. This indicates saturation in both lines combined
with an unobscured continuum. A rough estimate based
on visual inspection suggests an absorption covering factor of about 0.75. The EWs listed
in M06 must be corrected for this factor in order to compare with the calculations.

Regarding the gas density, in principal 
this can be directly obtained from the relative intensity of the 
Fe~{\sc xii} lines since one of the lines, at 11.92\AA, originates from an excited $2s^22p ^2P^0_{3/2}$ level at 14.6 eV 
above ground. The known collision strength and decay rate of this level indicate a critical 
density of order $10^{14}$ cm$^{-3}$ for 
the temperature found here ($\sim 2 \times 10^6$\,K).
However,  this is only true for optically thin transitions while for
the density estimated here (about $10^{13}$ cm$^{-3}$
at the base of the wind) and the assumed column density, the lines in question
are optically thick (the ground level transitions
are starting to saturate while the optical depth of the 11.92\AA\ line
is of order unity). This changes the line ratios in favor of the smallest optical depth line
and  results in a 
satisfactory agreement between the observations and the calculations of the EWs of
all Fe~{\sc xxii} lines.

The results of two models with different assumptions on the radial velocity of the wind,
$x=0$ and $x=0.3$, are summarized in Table 1.
The calculated spectrum for the $x=0$ atmosphere is shown in Fig. 1 on a scale similar to the
one shown in M06 (supplementary Fig. 1). The computed spectrum was convolved with the HEG
instrumental profile to enable a direct comparison with the observations.
The diagram shows the great similarity of the model with the \chandra\ observations\footnote{Note that the flux
scale used here  corresponds to the absorption corrected spectrum while M06 plot the observed spectrum. Note also
that the Fe~{\sc xxiv} L-edge at 6.06\AA\ looks artificially prominent because the current model does not include the high
order L-shell lines of this ion and the Ni~{\sc xxvi} line at 6.05\AA}.
\begin{table}
\caption{Parameters of thermal wind models.}
\begin{center}
\begin{tabular}{ccccccc}
\hline
    $\xi_0$   & $r_0$             &     $n_0$                &  column density          &x    & $V_{expan}$& EW(ob
s)/EW(model) \\
\hline
 $1.1 \times 10^3$           & $10^{10.73}$ cm   &  $10^{13.0}$ cm$^{-3}$  &  $10^{23.73}$ cm$^{-2}$  & 0   & 300
   &  0.99$\pm 0.40$   \\
$9.2 \times 10^2$ & $10^{10.65}$ cm   &  $10^{13.25}$ cm$^{-3}$  &  $10^{23.78}$ cm$^{-2}$  & 0.3 & 400        &  1.03
$\pm 0.46$   \\
\hline
\end{tabular}
\end{center}
\label{table:wind_models}
\end{table}

\begin{figure}
\plotone{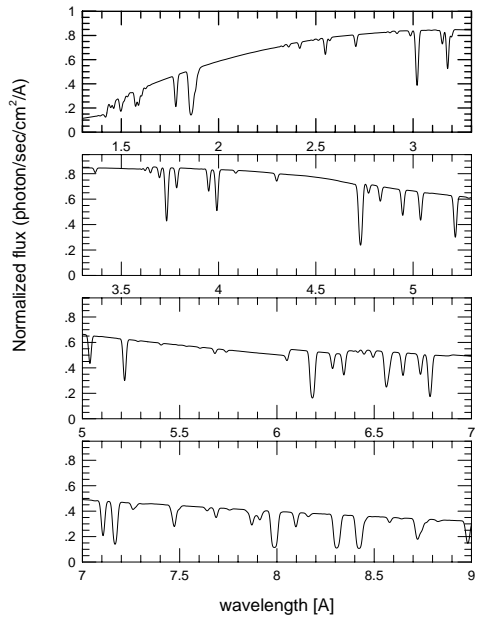}
\caption{
A thermal wind model for \GR\ (the absorption spectrum of model A).
 The continuum is corrected for galactic absorption and an absorption covering fraction
 of 0.75 is assumed.
}
\label{fig:wind_spectrum}
\end{figure}

 To make a more quantitative comparison I show, in Fig. 2, a comparison of 53 EWs of observed and calculated
saturated and unsaturated lines. These lines are basically the
 ones used by M06 in their Fig. 2 except that the present calculations do not include nickel and sodium lines.
 The values plotted are EW(observed)/EW(model) where EW(model) is obtained by assuming Gaussian
line profiles (damping winds are not important at such velocities) and where the computed EWs are corrected for the blending of several
line pairs.
The diagram shows a uniform distribution
with no obvious ionization trends.
The mean and the standard deviation of this ratio,
for the two models detailed in Table 1, are $0.99 \pm 0.40$ for the $x=0$ model and $1.03 \pm 0.46$ for the
 $x=0.3$ model. Both are considered satisfactory
given the measurement uncertainties
and the simplicity of the model and both are of comparable quality to the M06 fit (their Fig. 2).
There are only two notable deviations.
The first is Ne~{\sc ix}\,13.45\AA\ where
  EW(observed)/EW(model)=2.3$\pm 0.5$. This is suspected to be due to an overestimate of the
measured EW and can be tested by measuring a different Ne~{\sc ix} line observed in the spectrum but not listed
  in M06. The second is Fe~{\sc xxii}\,11.77\AA\ where EW(observed)/EW(model)=0.43$\pm 0.06$. However,
the spectrum shows several additional Fe~{\sc xxii} lines that were not identified by M06 (e.g. at 11.54 and 11.42\AA) that
seem to be in better agreement with the model, although their EWs was only estimated from the plot and hence
was not included in the comparison.
The ground-to-excited level line ratios in Fe~{\sc xxii} is in good agreement with the observations.
\begin{figure}
\plotone{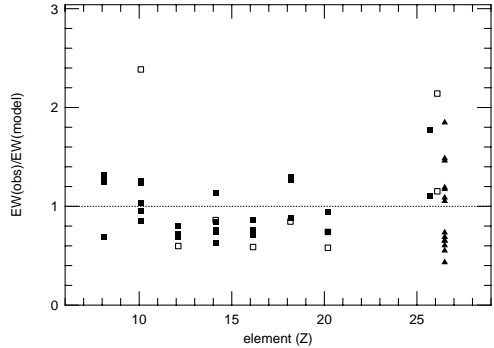}
\caption{Observed over calculated EWs for the absorption lines in model A.
Full squares represent
lines from H-like ions, empty squares are lines of He-like ions and triangles are lines of
 Fe~{\sc xxii},  Fe~{\sc xxiii} and Fe~{\sc xxiv}.
}
\label{fig:EWs}
\end{figure}

\section{Discussion}

The \chandra\ observations of \GR\ are perhaps the best
indication, so far, of X-ray heated flows in galactic
sources containing accretion disks. M06 have argued that the origin of this wind is magnetic because
two alternative mechanisms that can drive such flows are inefficient.
One possibility, acceleration by radiation pressure force, was ruled out on ground that the force multiplier
in such highly ionized gas is too small to drive the observed flow. The models presented here agree with
this conclusion.  The value of the
force multiplier ($\sim 1.5$) is more than an order
of magnitude too small to drive a wind given \Ledd\ of the source.
The second possibility, of an
X-ray heated wind, was rejected on the basis of the derived location of the absorbing gas
which was obtained by computing a detailed photoionization
model to fit the observed spectrum. This was found by M06 to be well inside the minimum distance where
thermally driven winds are thought to be important. 

The model presented here is of similar quality to the M06 model
yet the gas location is some two orders of magnitude further out. This is the region where thermal,
X-ray heated outflow are likely to be important.
While not a complete 2D time dependent wind model, all the properties found here, including the location,
 velocity and column density of the gas,
 are consistent with theoretical expectations. The mass loss rate in this model is
$\dot{m_w} \simeq 2 \times 10^{19} \Omega_{em}/4 \pi $ gr~s$^{-1}$ (note a comment below about
non-radial streamlines). For an accretion efficiency of 0.1,
this is about a factor of $50 \Omega_{em}/4 \pi$ larger than the accretion rate by the central object.
Given the small derived $\Omega_{em}$, this is
in general agreement
with the Begelman et al. (1983) results.

Realistic 2D numerical simulations that include the possibility of a non-uniform medium and  the disk-wind
interaction are required to fully explain the spectrum of \GR. Such calculations
must address several important issues that could not be included in the present model:
1. Thermal instabilities in the outflowing gas.
2. A multi-component patchy absorber.
3. Non-radial streamlines that will modify the simplified continuity condition (eqn. 3) and hence
also the radial dependence of $\xi$ (e.g. a decrease of $\xi$ with $r$).
4. Radial dependence of the covering factor.
5. Time dependent $L_X$ reflecting the observed variability of the source.
The simple 1D model presented here is already in good agreement
with the observations and is likely to be a major component of the
more advanced computations.

All the above  does not exclude the possibility that magnetically driven winds are important sources of driving
material near galactic and extragalactic BHs, including \GR. Such winds have been proposed and explored 
in numerous papers and provide a natural scenario for many observed systems (e.g. Blandford \& Payne 1982;
Hawley, Gammie \& Balbus, 1995). However, given the success of the simple thermal wind model to explain
 the observations of \GR, there seems to be no need to
 invoke a magnetically powered wind as a major contributer in this case.

\begin{acknowledgements}

I am grateful to the Kavli Institute at SLAC (KIPAC) for their hospitality during part of
my sabbatical.
I am grateful to Jon Miller and John Raymond for providing useful information about the \chandra\
observations and the model presented in M06. John Raymond has pointed out the importance of
the Fe~{\sc xxii} lines in deriving the gas density.
 I also thank Steve Kahn, Roger Blandford, Doron Chelouche, Ming-Feng Gu and Gary Ferland for useful discussions.

\end{acknowledgements}


%



\begin{thebibliography}{}

\bibitem[Badnell(2006)]{2006A&A...447..389B} Badnell, N.~R.\ 2006, \aap,
447, 389

\bibitem[Begelman et al.(1983)]{1983ApJ...271...70B} Begelman, M.~C.,
McKee, C.~F., \& Shields, G.~A.\ 1983, \apj, 271, 70

\bibitem[Blandford \& Payne(1982)]{1982MNRAS.199..883B} Blandford, R.~D.,
\& Payne, D.~G.\ 1982, \mnras, 199, 883

\bibitem[Chelouche \& Netzer(2005)]{2005ApJ...625...95C} Chelouche, D., \&
Netzer, H.\ 2005, \apj, 625, 95

\bibitem[Hawley et al.(1995)]{1995ApJ...440..742H} Hawley, J.~F., Gammie,
C.~F., \& Balbus, S.~A.\ 1995, \apj, 440, 742

\bibitem[Landi \& Gu(2006)]{2006ApJ...640.1171L} Landi, E., \& Gu, M.~F.\ 
2006, \apj, 640, 1171

\bibitem[Miller et al.(2004)]{2004ApJ...601..450M} Miller, J.~M., et al.\
2004, \apj, 601, 450

\bibitem[Miller et al.(2006)]{2006Natur.441..953M} Miller, J.~M., Raymond,
J., Fabian, A., Steeghs, D., Homan, J., Reynolds, C., van der Klis, M., \& 
Wijnands, R.\ 2006, \nat, 441, 953 (M06)

\bibitem[Netzer et al.(2003)]{2003ApJ...599..933N} Netzer, H., et al.\
2003, \apj, 599, 933
\bibitem[Netzer(2004)]{2004ApJ...604..551N} Netzer, H.\ 2004, \apj, 604,
551
\bibitem[Netzer et al.(2005)]{2005ApJ...629..739N} Netzer, H., Lemze, D.,
Kaspi, S., George, I.~M., Turner, T.~J., Lutz, D., Boller, T., \&
Chelouche, D.\ 2005, \apj, 629, 739

Netzer, H.\ 2005, \apj, 625, 95

\bibitem[Orosz \& Bailyn(1997)]{1997ApJ...477..876O} Orosz, J.~A., \& 
Bailyn, C.~D.\ 1997, \apj, 477, 876 

\bibitem[Woods et al.(1996)]{1996ApJ...461..767W} Woods, D.~T., Klein,
R.~I., Castor, J.~I., McKee, C.~F., \& Bell, J.~B.\ 1996, \apj, 461, 767 (W96)



 
\end{thebibliography}
\end{document}